\documentclass[fleqn]{wlscirep}
\newcommand{\av}[1]{\langle {#1} \rangle}
\usepackage{pdfpages}

\title{Griffiths phases and localization in hierarchical modular networks}

\author[1,*]{G\'eza \'Odor}
\author[2]{Ronald Dickman}
\author[3]{Gergely \'Odor}
\affil[1]{MTA-MFA-EK Research Institute for Technical Physics and Materials 
Science, H-1121 Budapest, P.O. Box 49, Hungary}
\affil[2]{Departamento de Fisica and National Institute of Science and
Technology of Complex Systems, ICEx, Universidade Federal de Minas Gerais, 
Caixa Postal 702, 30161-970, Belo Horizonte - Minas Gerais, Brazil}
\affil[3]{Massachusetts Institute of Technology, 77 Massachusetts Avenue
Cambridge, MA 02139-4307, USA}
\affil[*]{odor@mfa.kfki.hu}

\begin{abstract}

We study variants of hierarchical modular network models suggested by Kaiser 
and Hilgetag [{\it Front. in Neuroinform.,} {\bf 4} (2010) 8] to model 
functional brain connectivity, using extensive simulations and quenched 
mean-field theory (QMF), focusing on structures with a connection 
probability that decays exponentially with the level index.
Such networks can be embedded in two-dimensional Euclidean space.
We explore the dynamic behavior of the contact process (CP)
and threshold models on networks of this kind, including hierarchical trees.
While in the small-world networks
originally proposed to model brain connectivity,
the topological heterogeneities are not strong enough to induce deviations
from mean-field behavior, we show that a Griffiths phase can emerge under
reduced connection probabilities, approaching the percolation
threshold. In this case the topological dimension of the networks is finite,
and extended regions of bursty, power-law dynamics are observed. Localization
in the steady state is also shown via QMF.
We investigate the effects of link asymmetry and coupling disorder, and show that
localization can occur even in small-world networks with high
connectivity in case of link disorder.
\end{abstract}

\begin{document}

\maketitle

\section*{Introduction}

In neuroscience, the criticality hypothesis asserts that the brain is in 
a critical state, at the boundary between sustained activity and an 
inactive regime.
Theoretical and experimental studies show that critical systems
exhibit optimal computational properties, suggesting why criticality may
have been selected in the evolution of the nervous system \cite{LM07}.
Although criticality has been observed in cell cultures \cite{BP03,T10},
brain slices and anesthetized animals \cite{H10,R10}, the debate regarding
criticality in alert animals and humans continues \cite{B06,D12,P14}.
Thus the criticality hypothesis remains controversial
in brain science; for a review see \cite{BT12}.

Normally, for a system to be at criticality, certain control parameters 
need to be tuned precisely, raising the perennial
question of how such tuning is achieved in the absence of outside intervention.
The possibility of self-tuning is well known in statistical physics;
the paradigm of self-organized criticality (SOC) has been studied since
the pioneering work of \cite{Bak}.
Simple homogeneous models such as the stochastic sandpile exhibit
criticality with power laws both in statics and dynamics.  This has been
understood as the result of a control mechanism that forces the system to
an absorbing-state phase transition \cite{pruessner}

Real nervous systems, however, are highly inhomogeneous, so that one must take
into account the effects of heterogeneities. It is well known in
statistical physics that disorder can cause
rare-region (RR) effects \cite{Vojta} that smear the phase transitions.
These effects can make a discontinuous transition continuous,
or generate so-called Griffiths phases (GP) \cite{Griffiths},
in which critical-like power-law dynamics appears over an extended
region around the critical point.  In these regions, moreover,
non-Markovian, bursty behavior can emerge as a consequence of
a diverging correlation time. The inter-communication times of
the nodes (which possess no internal memory) follow a fat-tailed 
distribution \cite{burstcikk}.
Thus, heterogeneities widen the critical region and can contribute
to the occurrence of power laws.  This provides an alternative mechanism for
critical-like behavior without fine tuning, although attaining the GP does
require some rough tuning, of the sort that is not difficult to find
in biological systems.

It was shown recently that the topological heterogeneity of the underlying
networks can result in GPs in finite-dimensional systems
\cite{GPCNlong} and can be a reason for 
the working memory in the brain \cite{Johnson}.
Although many networks exhibit the small-world property and so have
an infinite topological dimension, naturally occurring networks are
always finite, exhibit cutoffs, and therefore GPs can be expected as a
consequence of inhomogeneous topology \cite{basiscikk}.

Many real networks can be partitioned seamlessly into a collection
of modules. Each module is expected to perform an identifiable task,
separate from the function of others \cite{NR10}.
It is believed that the brain is organized hierarchically
into modular networks across many scales, starting from cellular
circuits and cortical columns via nuclei or cortical areas
to large-scale units such as visual or sensory-motor cortex.
At each level, connections within modules are denser than between
different modules \cite{SCKH04,Sporns,Kaiser,MLR10}.
Although empirical data confirm this modular organization on some scales
\cite{CCH10}, the detailed organization of brain networks is not yet
experimentally accessible.

Two particular kinds of hierarchical modular networks (HMN-1,HMN-2) 
model were proposed and investigated numerically and analytically in 
\cite{MMNat}. 
On large-world HMNs, which imply a finite topological
dimension $D$, models of the spread of activity exhibit power-law
dynamics and rare region effects. However, these power laws are
system-size dependent, so that true GP behavior has not yet been proven.
These authors also simulated spreading on empirical brain networks, 
such as the human connectome and the nervous system of {\it C. elegans}.
Available empirical networks are much smaller than the synthetic
ones and deviations from power laws are clearly visible.
Both anatomical connections \cite{SCKH04} and the synchronization networks
of cortical neurons \cite{22} suggest small-world topology \cite{HGP06}.
The brain network modules of \cite{MMNat} are weakly coupled in 
such a way that these HMNs are near the percolation threshold,
as in the case of the models introduced in \cite{GPCNlong}. Note, however that
requiring the network to be near percolation again raises tuning and
robustness issues. Having weaker ties would lead to fragmented networks,
while stronger ties result in infinite $D$ and the absence of a GP.
In the present work we do not assume such fine tuning: we maintain a
high density of short edges, rendering the networks well connected.
Another way of preserving the integrity of HMN networks with finite
dimension involves the random tree-like structures studied here.

To study synchronization \cite{Sync}, commonly expected in brain dynamics,
the Kuramoto model \cite{Kur} has been implemented \cite{Matj} on the 
same networks as studied in \cite{MMNat}.
In this case even weaker rare-region effects were found,
resulting in stretched exponential dynamics.
One of the main purposes of our study is to delineate conditions a GP.
While the networks studied are of finite size,
repeating the process on many network realizations and averaging over them,
we clearly see convergence towards GP dynamics. We assume that
multiple random network realizations may occur in the brain over time,
due to reconfigurations of the synapses or as a consequence of
weakly coupled sub-modules and changes in overall intensity of 
connections between different brain regions due to neuromodulation.

In addition to the dimension, other topological features,
which have yet to be studied systematically, may also influence
RR effects. The clustering coefficient, defined as
\begin{equation}
C(N) = 1/N \sum_i \frac{2n_i}{k_i(k_i-1)} \ ,
\end{equation}
where $n_i$ denotes the number of direct links interconnecting the $k_i$ nearest
neighbors of node $i$, exhibits different scaling behavior in modular networks
than in unstructured networks.  While in SF networks it decays as
$C(N) \sim N^{-3/4}$ and in random networks as $C(N) \sim N^{-1}$,
in HMNs it is constant.
Furthermore, while in SF networks the clustering is degree-number independent,
in HMNs it decays as $C(k) \sim k^{-\beta}$,
with $\beta$ in the range 0.75 - 1 \cite{RB03}.
This means that the higher the node degree, the smaller is its
clustering coefficient, possibly reducing its infectiousness
in an epidemic process. This suggests an enhancement of localization
as in SF models with disassortative weighting schemes \cite{BAGPcikk}.

In this work we study spreading models defined on HMNs embedded
in 2D space, and investigate how various inhomogeneities
(topological or intrinsic) contribute to the occurrence of
localization and GPs. We begin by considering purely topological
heterogeneity; later we study cases with disorder in the
connections as well.  It is well known that intrinsic link disorder
can appear as the consequence of asymmetry, connecting nodes
in the cortex (see \cite{Nep}). It has also been demonstrated that
weights can vary over 4-6 orders of magnitude; most
(including the majority of long-range connections) are actually
quite weak \cite{Markov}.

The relation of RR effects to localization in the steady
state has been studied recently \cite{basiscikk}.
Here we discuss localization results obtained
via a quenched mean-field (QMF) approximation.

In addition to serving as models of brain connectivity, hierarchical modular
networks are abundant in nature. They occur in social relations \cite{hsoc},
in the WWW \cite{hwww}, metabolic \cite{hmeta} and information systems
\cite{hinter} for example.  Thus the results reported here should find
application in these and related areas.

\section*{Hierarchical modular networks}

Recent studies indicate that activity in brain networks persists between
the extremes of rapid extinction and global excitation.
This so-called limited sustained activity (LSA) does not occur in
spreading models defined on structureless, small-world networks.
On the other hand, brain networks are known to exhibit hierarchical modular
structure.  This motivated Kaiser and Hilgetag (KH) to perform numerical
studies on such networks, to investigate topological effects on LSA \cite{KH}.
Their hierarchical model reflects general features of brain connectivity
on large and mesoscopic scales.  Nodes in the model were intended to
represent cortical columns rather than individual neurons.
The connections between them were taken as excitatory, since there
appear to be no long-distance inhibitory connections within
the cerebral cortex \cite{LatNir2004}.

Kaiser and Hilgetag generated networks beginning at the highest level, and adding
modules to the next lower level, with random connectivity within
modules. They explored hierarchical networks with different numbers
of levels, and numbers of sub-modules at each level.
The average degree (over all nodes in the network) was set to 
$\langle k\rangle = 50$, motivated by experimental studies.
They investigated different topologies by varying the edge density
across the levels. All the networks studied by KH are of small-world type, i.e.,
they have an infinite topological dimension.

The spreading model investigated by KH is a two-state threshold model,
in which, at each time step, inactive nodes are activated with probability
$\lambda$ if at least $m$ of their neighbors are currently active,
and active nodes deactivate spontaneously with probability $\nu$.  This
model is very similar to reaction-diffusion models known in statistical
physics \cite{DickMar,rmp,odorbook}, with a synchronous cellular automaton 
(SCA) updates.
Starting from an initial state in which a localized set of nodes is active,
these authors followed the density of active sites and their
localization up to $t\simeq 200$ time steps on networks of sizes $N \le 512$.

Kaiser and Hilgetag found that LSA can be found in a larger
parameter range in HMNs, as compared with random and non-hierarchical
small-world networks. The optimal range of LSA was found in networks
in which the edge density increased from the top level ($l=l_{max}$)
to the bottom ($l=l_1$).
Such topologies foster activity localization and rare-region effects.

In this paper we investigate HMNs which possess increasing
edge density from top to bottom levels, as did KH, but with {\it finite}
topological dimension.
We will show that although localization is seen in small world
networks, to observe GPs, with power-law dynamics, we need
networks of finite topological dimension.

\begin{figure}[h]
\includegraphics[height=5.5cm]{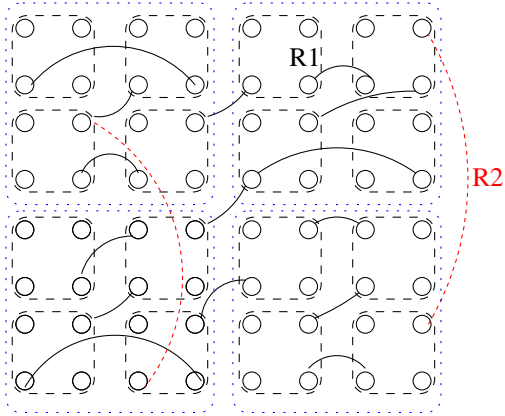}
\caption{\label{mod} Two lowest levels of the HMN2d hierarchical network 
construction.
Dashed lines frame bottom level nodes, which are fully connected there,
dotted lines frame the nodes of the next level. 
The solid lines denoted R1 are randomly chosen connections among the
bottom level modules, ensuring single connectedness of the network,
while those denoted R2 provide random 
connections on the next level. Links can be directed.}
\end{figure}

To generate a hierarchical modular network, we define $l_{max}$ levels on
the same set of $N=4^{l_{max}}$ nodes; on the $l^{th}$ level we
define $4^l$ modules. We achieve this by splitting each module into four
equal sized modules on the next level, as if they were embedded in a regular,
two-dimensional (2d) lattice (see Figure~\ref{mod}).
The probability $p_l$ that two nodes are connected by an edge follows
$p_l \simeq b 2^{-s l}$ as in \cite{KH}, where $l$ is the greatest
integer such that the two nodes are in the same module on level $l$.
After selecting the number of levels and nodes we fix the average node
degree, $b = \langle k\rangle / 2$, and generate the
adjacency matrix $A$ by proceeding from the highest to the lowest level.
We fill the submatrices with zeros and ones and copy them to appropriate
diagonal locations of $A$.
We allow unidirectional connections, which is
more realistic.  In fact, disorder associated with the link orientations turns
out to be necessary to observe GPs. Finally, we connect the lowest-level modules
with an edge, chosen randomly from nodes of $l_1$.
This provides a short-linked base lattice that guarantees connectivity.

In a preliminary study, the extra edges, corresponding
to the base lattice, were not added. The resulting networks
typically consist of a large number of isolated connected components;
the GP effects observed in these structures are a consequence of 
fragmented domains of limited size.
Measurements of axon-length distributions in several real neural networks
show a large peak at short distances followed by a long flat tail.
Thus there is a dense set of local edges in addition to a sparse network 
of long-range connections, best fit by an exponential function \cite{Alen}.
Typical estimates indicate that $\simeq 90 \%$ of all cortical
connections are formed at the local level
(i.e., within a radius of $0.5$mm); only the remainder leave the local gray
matter and project to remote targets. Therefore, we also investigate a
variant of HMN2d networks in which the lowest-level modules are fully 
connected,
and there is a single link among the nodes of modules on level $l_2$.
To broaden further the range of structures, we study
{\it hierarchical modular trees} (HMTs), which possess the minimum number
of edges required for connectivity, and so have no loops.  Construction of HMTs is
described in the Supplementary material.

\subsection*{Relation to Benjamini-Berger networks}

Due to the embedding, there is  a correspondence with Benjamini-Berger
(BB) networks \cite{BB}.
BB networks are generated on Euclidean lattices,
with short links connecting nearest neighbors. In addition, the network contains
long links, whose probability decays algebraically with
Euclidean distance $R$:
\begin{equation} \label{BB}
p(R) \sim R^{-s} \ .
\end{equation}
Here we consider modified BB networks, in which the long links are
added level by level, from top to bottom, as in \cite{KH}.
The levels : $l=1,...,l_{max}$ are numbered from the bottom to top.
The size of domains, i.e., the number of nodes in a level, grows as $N_l = 4^l$
in the 4-module construction, related to a tiling of the
two dimensional base lattice. Due to the approximate distance-level relation,
$R\simeq 2^l$, the long-link connection probability on level $l$ is:
\begin{equation}
p_l = b \left( \frac{1}{2^s} \right)^l \ .
\end{equation}
Here $b$ is related to the average degree of a node $\langle k \rangle$.

It is known that in a one-dimensional base lattice the BB construction results
in a marginal case, $s=2$, in which the topological dimension is finite and
changes continuously with $b$.
For $s < 2$ we have small world networks, while for $s > 2$ the
the topological dimension is the same as the base lattice ($d=1$) in BB
networks.
The HMN-1 networks studied by Moretti and Mu\~noz~\cite{MMNat} are similar to
the BB model, without the 1d base lattice, but with the inclusion of a
HMN topology.
These authors simulated spreading models on HMN-1 with finite topological
dimension at $p_l \sim (\frac{1}{4})^l$ and found GPs.
Given the above mapping, this result is not very surprising, because this HMN
corresponds to the $s=2$ case, staying close to the percolation threshold
of the network. To ensure connectivity, Moretti and Mu\~noz added extra
links to the large connected components \cite{MMNat}; they assert that
the topological dimension remains finite, despite the additional links.
In networks with finite sizes this is certainly true, however in the infinite
size limit such a system is also infinite dimensional, so that we don't expect
true GPs. 
For the HMN-2, the number of connections between blocks at every level
is a priori set to a constant value. In this case numerics of 
\cite{MMNat,round2} suggest GP effects.

Here we embed the HMNs in 2d base lattices, which is closer
to the real topology of cortical networks.
In this case the threshold for marginality
(i.e., continuously changing dimension and exponents) is expected to be at
$s=4$. We confirm this by explicit measurements of the topological dimension.
Furthermore, we study two-dimensional HMNs with different $s$ values and find
GP effects for finite topological dimensions, both at the percolation 
threshold (for $s=3$), and for $s=6$, where the tail distribution of 
the link lengths decays very rapidly, in agreement with empirical
results for axon lengths.
  
In addition to the CP ($m=1$), we also consider threshold models ($m \ge 2$), 
which are expected to be more realistic for neuronal systems.

\subsection*{Dimension measurements}

To measure the dimension of the network we first compute the level structure
by running a breadth-first search from several, randomly selected seeds.
We count the number of nodes $N(r)$ with chemical distance $r$ or less
from the seeds and calculate averages over the trials.
We determined the dimension of the network from the scaling relation
$N \sim r^d$, by fitting a power-law to the data for $N(r)$.

\begin{figure}[h]
\includegraphics[height=5.5cm]{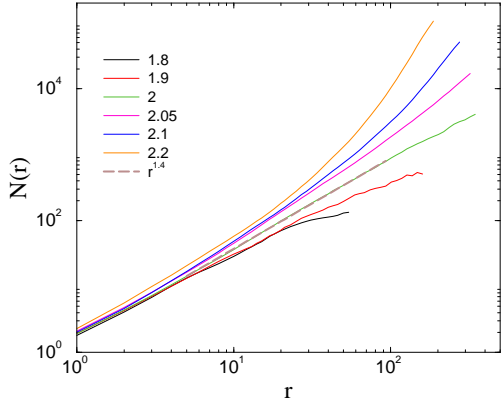}
\caption{\label{dim-s3}Number of nodes within the chemical distance $r$
in the $s=3$ HMN2d models with $l_{max}=9$. Different curves correspond to
different parameters: $b=\langle k \rangle/2$ as indicated.
The dashed line shows a power-law fit for $b=2$, with $N(r) \sim r^{1.4}$, 
suggesting a topological dimension of $d = 1.4$ at the percolation threshold. }
\end{figure}

\begin{figure}[h]
\includegraphics[height=5.5cm]{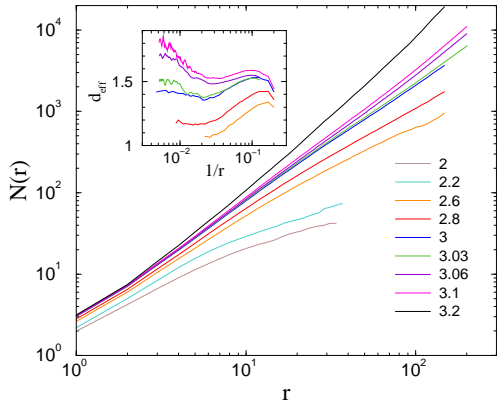}
\caption{\label{dim-s4}Number of nodes within chemical distance $r$
in HMN2d networks with $s=4$ and $l=9$ levels. Different curves correspond to
different $b=\langle k \rangle/2$ values as indicated. 
Inset: local slopes $d_{eff}$ of the $N(r)$ curves, 
defined in equation.~\ref{deff}.}
\end{figure}

At $s=3$ we observe a percolation threshold near $\langle k \rangle=4$,
for which the  topological dimension is finite: $d \simeq 1.4$
(see Figure~\ref{dim-s3}).
Note that the curves with large $\langle k \rangle$ exhibit
saturation, corresponding to a finite-size effect.
The case $s=4$ appears to correspond to a marginal dimension, for which,
above the percolation threshold $\langle k \rangle > 6$, one observes
a continuously varying topological dimension (see Figure~\ref{dim-s4}).
A more detailed, local slopes analysis via the
effective topological dimension

\begin{equation}  
\label{deff}
d_{\rm eff} = \frac {\ln[N(r)) / N(r')]}
      {\ln(r / r^{\prime})} \ ,
\end{equation}
\vspace{.1em}

\noindent is shown in the inset of Figure~\ref{dim-s4}, where
$r$ and $r'$ are neighboring measurement points. Saturation
to $s$-dependent dimensions can be read off the plot for $r>100$. 

\begin{figure}[h]
\includegraphics[height=5.5cm]{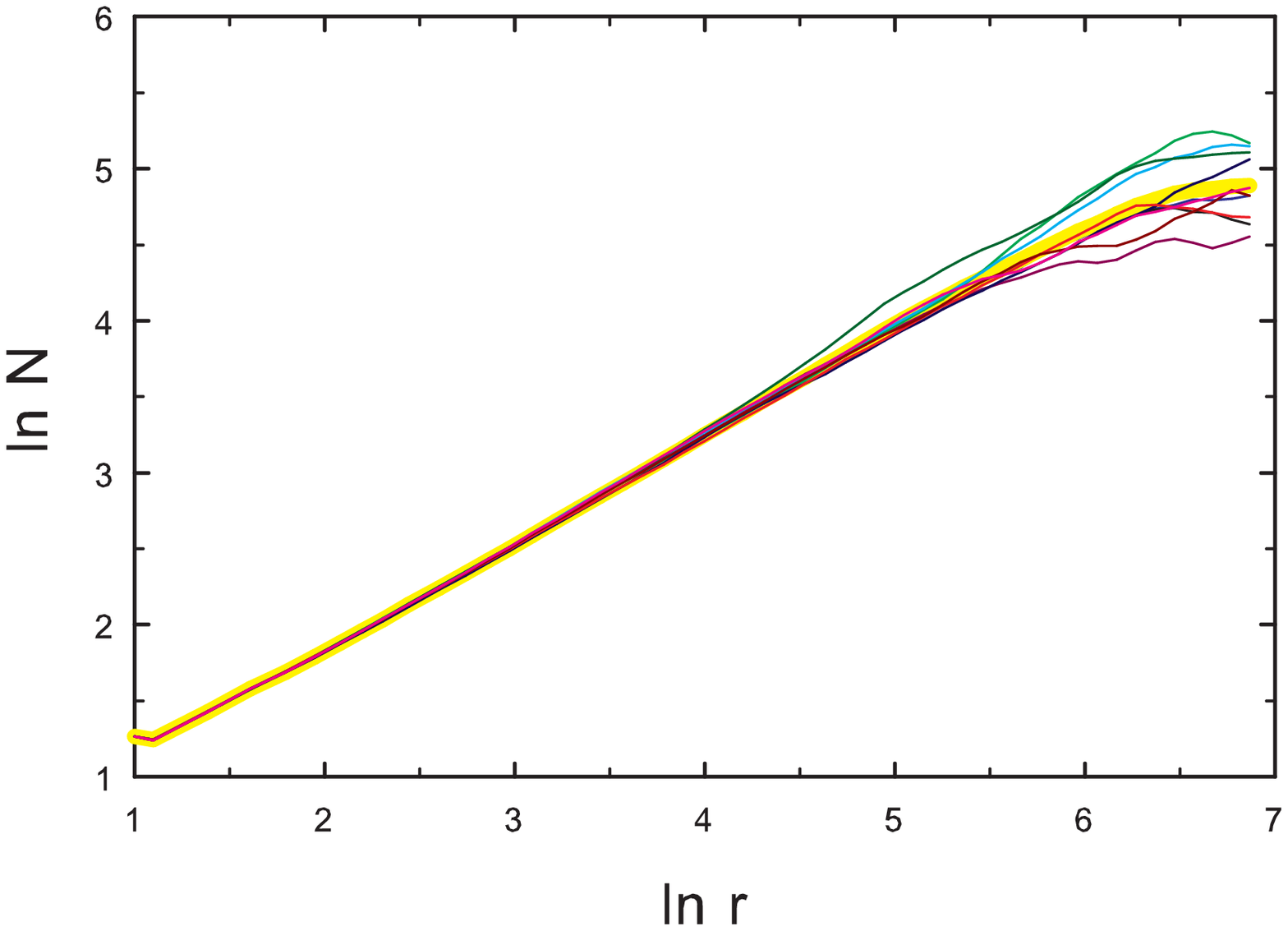}
\caption{\label{tentrees} Number of nodes within chemical 
distance $r$ in a set of ten random hierarchical trees with 262144 nodes
(thin curves). 
The broad yellow curve is an average over the set of ten structures.}
\end{figure}

\begin{figure}[h]
\includegraphics[height=5.5cm]{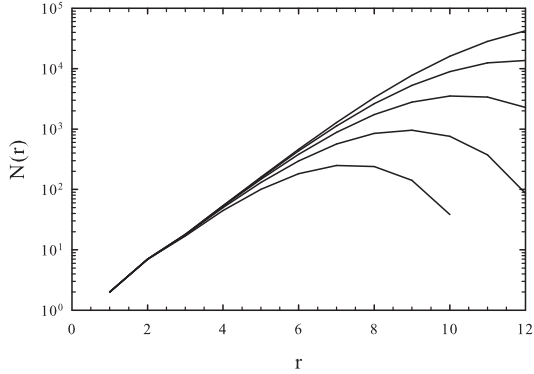}
\caption{\label{nrregtree} Number of nodes within chemical 
distance $r$ in regular hierarchical trees with 1024, 4096, ..., 262144 nodes
(lower to upper curves).}
\end{figure}

Random hierarchical trees also exhibit a finite topological dimension.  
Figurere \ref{tentrees} shows $N(r)$ for a set of ten independent
random trees.  The average of $N(r)$ over the set of ten trees follows 
a power law to good approximation, with $d \simeq 0.72$.  
(Note that in this case $N(r)$ is an average over {\it all} nodes.)  
For regular trees, by contrast, $N(r)$ grows exponentially with $r$, 
as shown in Figure~\ref{nrregtree}. Thus the regular tree construction 
results in a structure with infinite topological dimension.

\section*{Dynamic simulations}

\subsection*{Contact process}

The {\it contact process} (CP) \cite{harris74,liggett1985ips}, 
a Markov process defined on a network, can be interpreted as
a simple model of information spreading in social \cite{pv04},
epidemic spreading \cite{newref1,newref2,newref3}, or in brain networks 
\cite{MMNat}.
In the CP sites can be active or inactive.
Active sites propagate the activity to their neighbors at
rate $\lambda/k$, or spontaneously deactivate with rate
$\nu=1$ (for simplicity).

\begin{figure}[h]
\includegraphics[height=5.5cm]{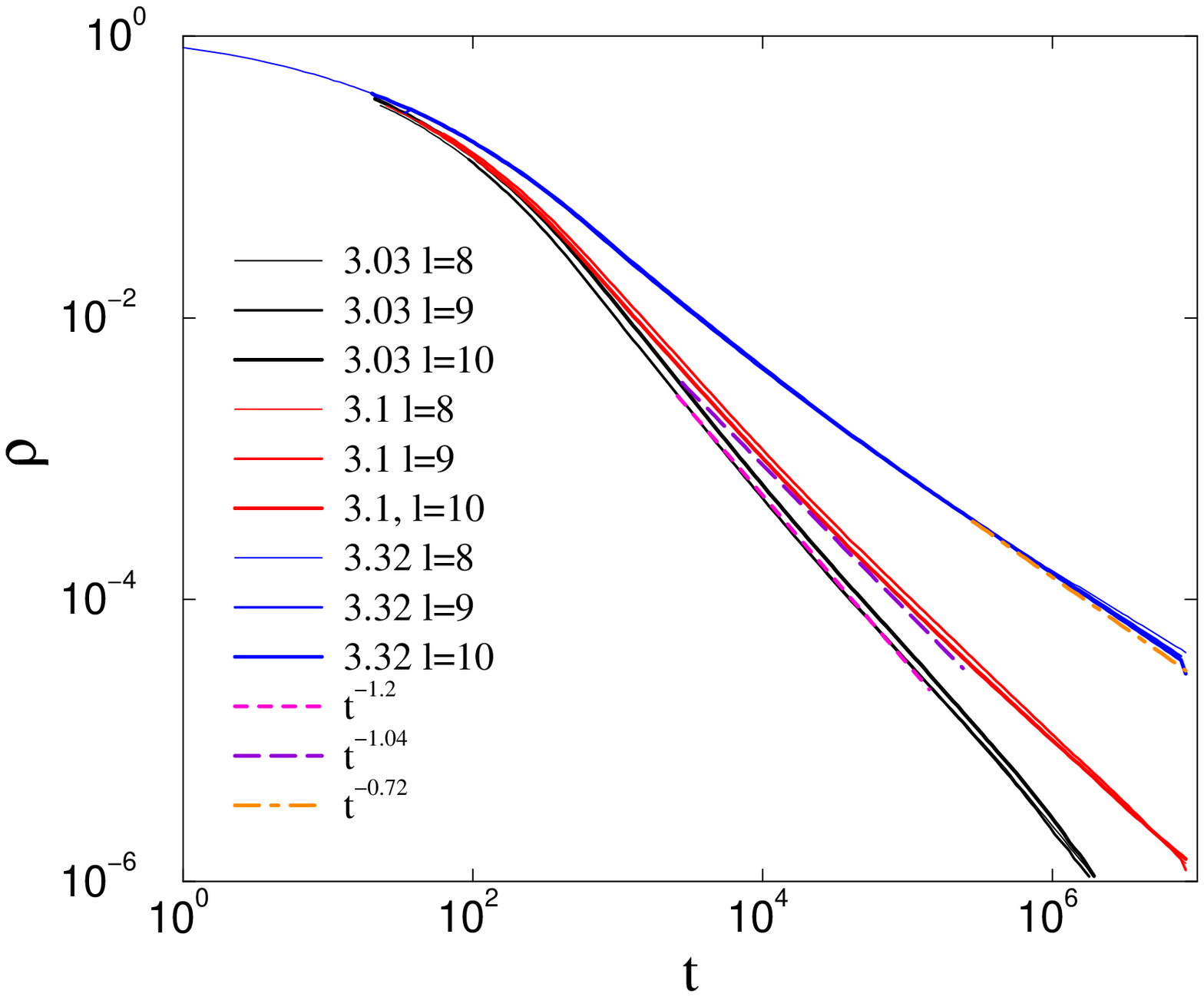}
\caption{\label{fss-s4} CP on asymmetric HMN2d networks with $s=4$:
decay of activity for $\lambda$ values as indicated.   System sizes $l_{max}=8,9,10$
(thin, medium, and thick lines, respectively).
Size-independent power laws are evidence of a Griffiths phase.}
\end{figure}

We perform extensive activity-decay simulations for HMN2d models with
$2b = \langle k\rangle=4$ and maximum levels: $l_{max}=8,9,10$.
In these networks we use directed links between nodes, similar to real nervous systems.
We follow $\rho(t)$ in $10-100$ runs for each independent
random network sample, starting from fully active initial states,
and averaging $\rho(t)$ over thousands of independent random
network samples for each $\lambda$.
In the marginal long-link decay case at $s=4$, a clear
GP behavior is found (see Figure~\ref{fss-s4}).

\begin{figure}[h]
\includegraphics[height=5.5cm]{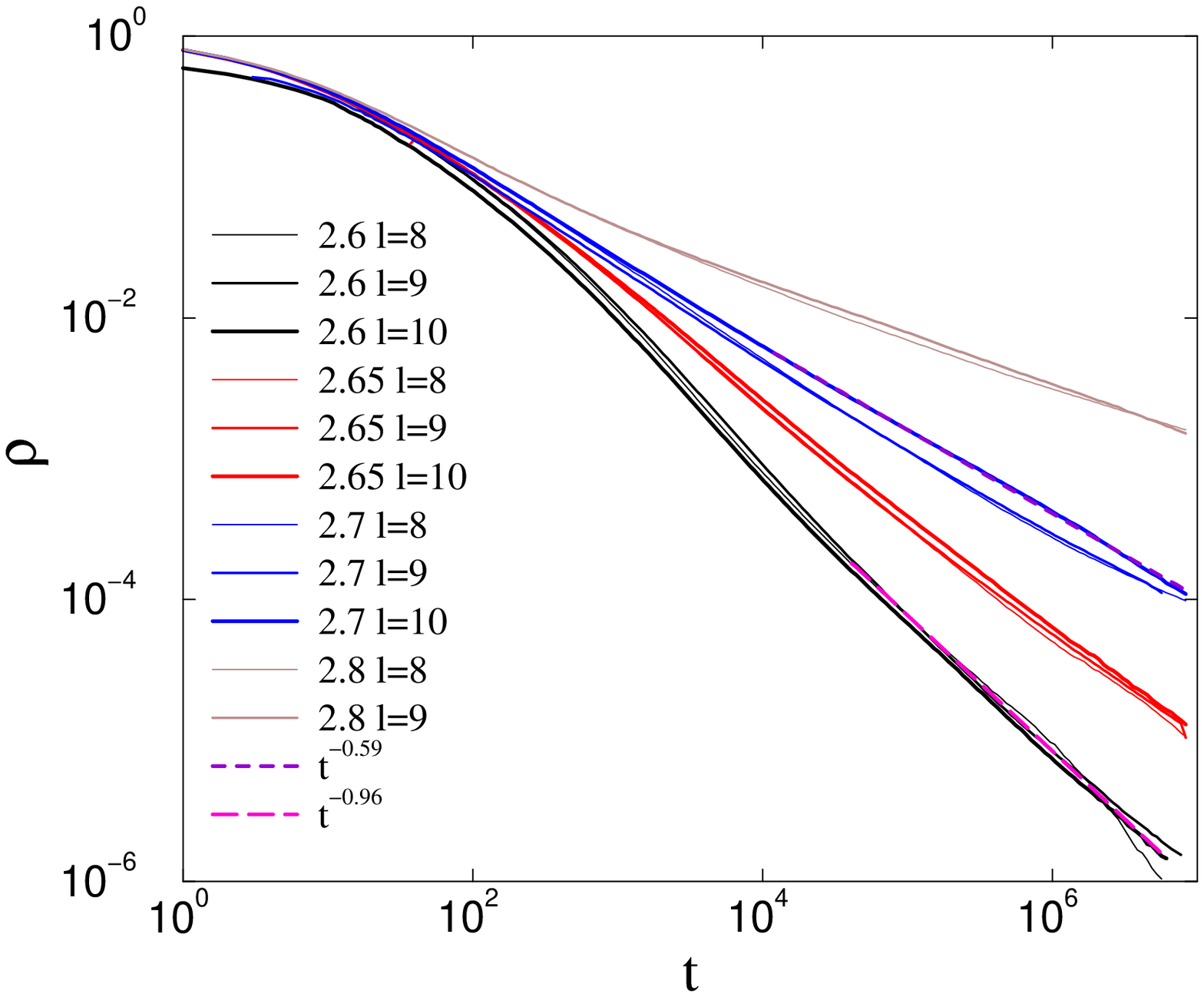}
\caption{\label{fss-s3} CP on asymmetric HMN2d networks
as in Figure~\ref{fss-s4}, but for $s=3$.}
\end{figure}

We show that a GP can occur for more general parameters than those studied in
\cite{MMNat}, by following the density decay in networks with $s = 3$
and $\langle k\rangle=4$ (i.e., at the percolation threshold).
Size-independent, nonuniversal power laws can be seen in Figure~\ref{fss-s3}.

When we increase the average degree $\langle k\rangle$,
the GP shrinks to a smaller range of $\lambda$ values, as in \cite{GPCNlong},
tending toward a simple critical phase transition. However, it is hard to
determine at precisely which value of $\langle k\rangle$ this happens.
We note that the connectivity of the network is not assured,
without the addition of extra links. Strictly speaking, however,
with this extension $s\le 4$ networks become infinite-dimensional in the
$N\to\infty$ limit.

\begin{figure}[h]
\includegraphics[height=5.5cm]{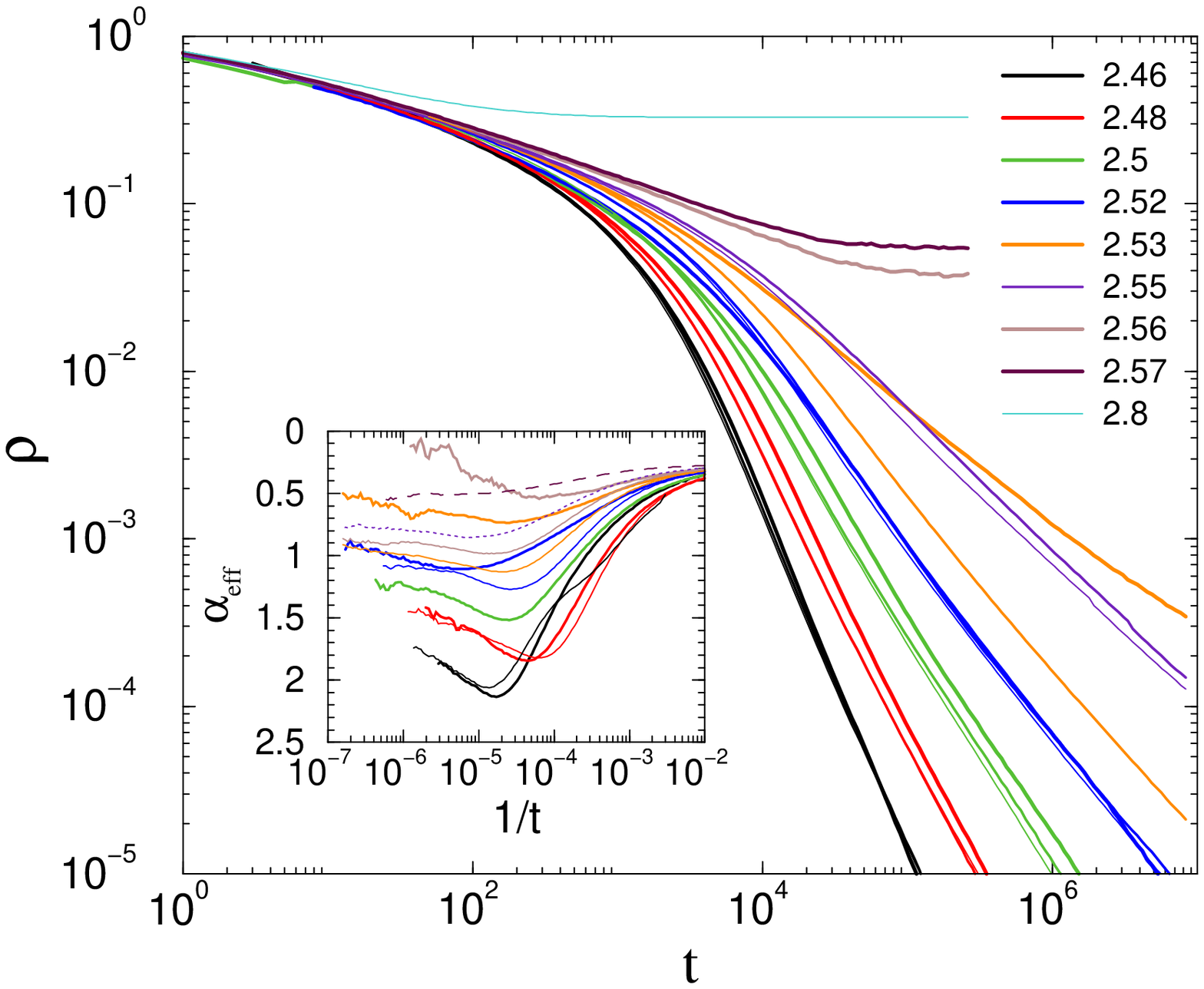}
\caption{\label{fss-s6} CP on asymmetric HMN2d networks with 
$s=6$ and $\langle k\rangle \simeq 4$:
decay of activity for $\lambda$ values as indicated.
System sizes as in Figure~\ref{fss-s4}.
Size-independent power laws are
observed for $2.45 < \lambda < 2.53$. Inset: local slopes
of the curves in the main plot. In the GP $\alpha_{eff}$ tends to
nonuniversal values with logarithmic corrections.
At $\lambda_c=2.53(1)$ we observe convergence to $\alpha_{eff}=1$; above
this threshold the effective exponents of larger systems veer upwards toward
zero as $t$ tends to infinity.}
\end{figure}

More importantly, GPs are also found in networks connected on the base level 
via short edges. We study the activity decay for $s=6$, which corresponds 
to fast decaying tail distribution for the long links, preserving finite 
topological dimension $d$.
In this construction the average degree at the bottom level is
$\langle k\rangle \simeq 6$, decreasing systematically with $l$, 
so that $\langle k\rangle \simeq 1$ for $l_{max}$.
Note that that the ratio of short to long links is $\sim 0.11$,
in agreement with results for real
neural networks \cite{Alen}.
Simulations again yield size-independent power-law decay curves,
confirming GP behavior, as shown in Figure~\ref{fss-s6}.

We also determined the effective decay exponent in the usual manner
(see \cite{odorbook}), via
\begin{equation}  \label{aeff}
\alpha_{\rm eff}(t) = - \frac {\ln[\rho(t)/\rho(t')]}
      {\ln(t/t^{\prime})} \ ,
\end{equation}
where $\rho(t)$ and $\rho(t')$ are neighboring data points.
The critical point can be located at $\lambda_c=2.53(1)$, showing
mean-field scaling: $\rho(t) \sim t^{-1}$.
Above this threshold power-laws can still be seen for smaller sizes
($l_{max}=8,9$), up to $\lambda \simeq 2.55$, but corrections to scaling
become stronger; the curves for $l_{max}=10$ exhibit saturation.
This is surprising, because in a disordered system with short ranged
interactions one would expect a critical phase transition with an
ultra-slow, logarithmic scaling. However, recent studies of the CP
in higher dimensions find mean-field criticality and GP
\cite{Vojta5d}. Our result suggests that the topological
heterogeneity generated by the long edges is not strong enough to
induce an infinite-randomness fixed point. Otherwise we
must assume very strong corrections to scaling in this case.

Disorder due to the randomly chosen orientations of the links turns out
to be relevant. For {\it symmetric} links our simulations yield nonuniversal
power laws, which appear to be sensitive to the system size,
suggesting the lack of a true GP in the infinite-size limit.

\subsection*{Burstyness in the CP}

\begin{figure}[h]
\includegraphics[height=5.5cm]{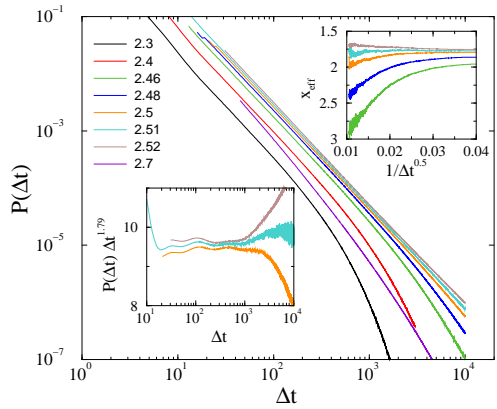}
\caption{\label{burst-s6} CP on asymmetric HMN2d networks
with $s=6$. Main plot : probability distribution, $P(\Delta t)$, of
inter-event times for $\lambda$ values as indicated; system size
$l_{max}=10$.
Power-law tails are evident for $2.46 < \lambda < 2.6$, with continuously 
changing exponents. 
Right inset: local slopes of the same curves, defined similarly as in
equation (\ref{aeff}) for: $\lambda=2.46, 2.48, 2.5, 2.52, 2.7$.
Left inset: $P^*(\Delta t) \equiv (\Delta t)^{1.79} P(\Delta t)$
for $\lambda=2.5, 2.51, 2.52$.}
\end{figure}

We study the distribution of inter-event times in the CP on
asymmetric, HMN2d networks with $s=6$, in a manner similar to that 
described in \cite{burstcikk}.
Starting from fully active states, on networks of size  $l_{max}=10$,
we measured $\langle \Delta t\rangle$ between successive activation
attempts. We followed the evolution up to $t_{max}=2^{18}$ Monte Carlo
steps (MCs), averaging over $1000-2000$ independent random networks with 
$10-100$ runs for each. Througout this paper time is measured by MCs.
As Figure~\ref{burst-s6} shows, fat-tailed distributions, 
$P(\Delta t) \sim (\Delta t)^{-x}$, emerge in the GP
for $2.46 < \lambda < 2.6$, while $P(\Delta t)$ decays exponentially
outside this region. The slopes of the decay curves are
determined via least-squares fits in the window $20 < \Delta t < 7000$.
For $\lambda =2.5$ we find $x=1.753(4)$, while
for $\lambda =2.52$ the exponent is slightly larger: $x=1.81(1)$.
These values are close to the auto-correlation exponent of the critical
$1+1$ dimensional CP as in \cite{burstcikk}, but exhibit deviations due to the
heterogeneities. This is understandable, since the effective dimension of
this HMN2d is close to one.
The scaling variable $P^*(\Delta t) \equiv (\Delta t)^{1.79} P(\Delta t)$ exhibits
log-periodic oscillations. This is the consequence of
the modular structure of the network. Furthermore, as in other GP models,
logarithmic corrections to scaling are expected.

We expect similar nonuniversal, control parameter dependent tails in
$P(\Delta t)$ in the GPs exhibited by the other HMN2d networks.
Furthermore, as shown in \cite{burstcikk}, power-law distributions
should arise for other initial conditions, such as localized activity.
This suggests that bursty inter-communication events in brain
dynamics arise spontaneously near the critical point, in the GP.

\subsection*{CP on random hierarchical trees}

We simulate the CP with {\it symmetric} links on random hierarchical trees 
(RHTs) of 262$\,$144 nodes.
We first perform quasistationary (QS) simulations \cite{qssim}, of the CP 
on a {\it single} RHT.  For one structure this yields 
$\lambda_c \simeq 2.72$; for another, independently generated structure 
of the same kind, we find $\lambda_c \simeq 2.76$.

\begin{figure}[h]
\includegraphics[height=6.5cm]{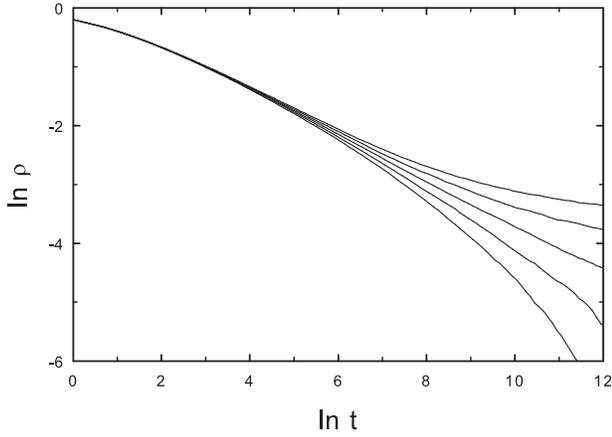}
\caption{\label{idtree11} CP on a random hierarchical tree:
activity versus time for $\lambda = 2.73$, $2.74$, ..., $2.77$ (lower to upper); 
system size: $N=262144$.
Power-law decay is evident only for $\lambda = 2.75$,
close to the estimate $\lambda_c = 2.76$ derived from QS simulations.
Each curve represents an average over $100$ realizations; all realizations are performed
on the same network.
}
\end{figure}

Our principal interest is in the decay of activity starting from all nodes active.
The decay of activity on a {\it single} RHT appears to follow the scenario familiar from the
CP on regular lattices: power-law decay is observed at $\lambda_c$ but not at nearby values,
as illustrated in Figure~\ref{idtree11}.  The randomness associated with a single RHT appears to be
insufficient to generate a Griffiths phase.

\begin{figure}[h]
\includegraphics[height=13cm]{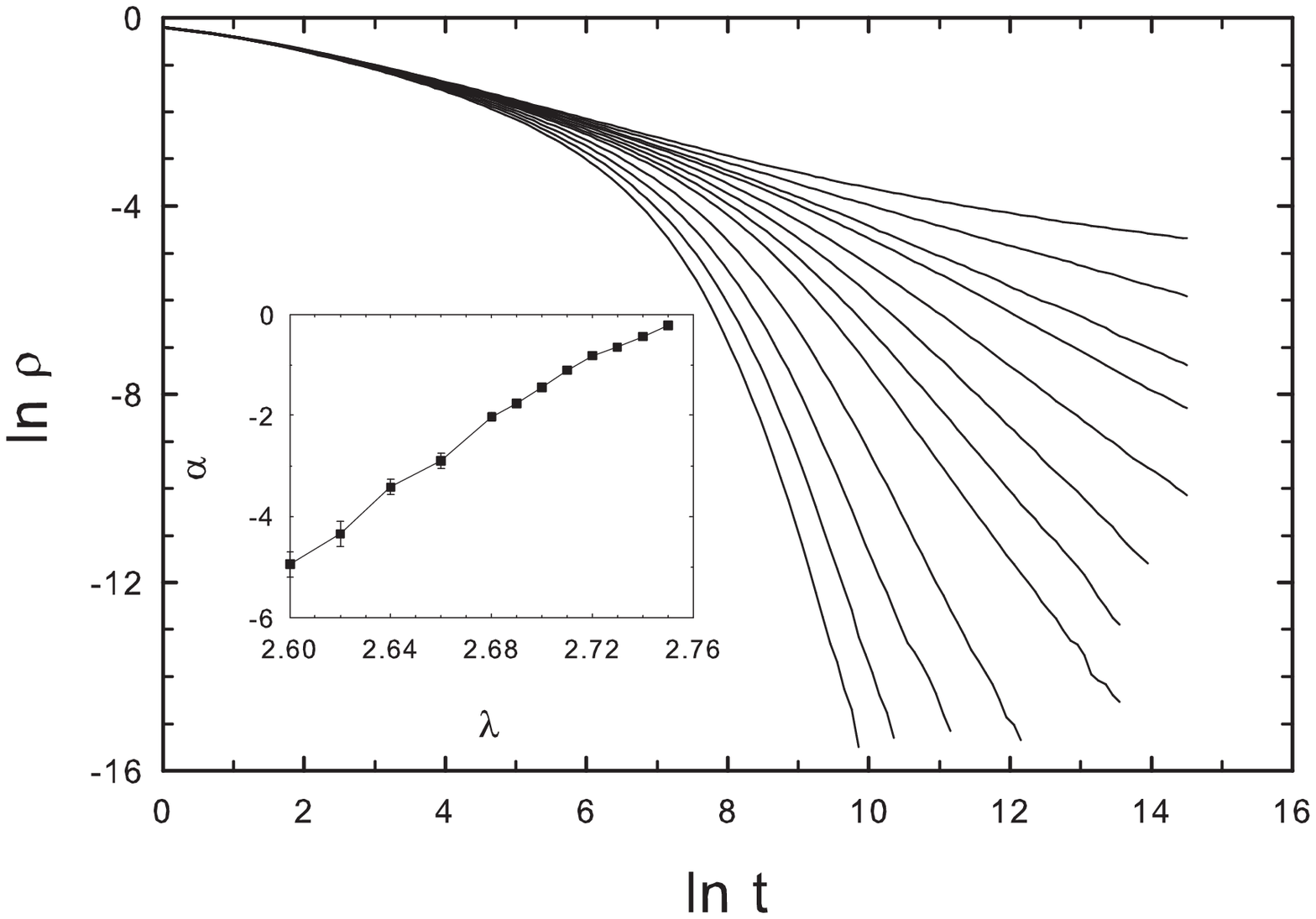}
\vspace{-3cm}

\caption{\label{rhtallb} CP on a random hierarchical trees:
activity versus time for $\lambda = 2.60$, $2.62$, $2.64$, $2.68$, $2.69$,
..., $2.75$ (lower to upper); system size $N=262144$.
Each curve represents an average over $10^3$ - $10^4$ realizations, each performed
on a different network.  Inset: decay exponent $-\alpha$ versus $\lambda$.}
\end{figure}

By contrast, evidence of a GP is found if we
average over many RHTs.  The activity averaged over a large set ($10^3$ - $10^4$) of independent
realizations, each on a different RHT, shown in Figure~\ref{rhtallb}, decays asymptotically as a power-law
over a fairly broad range of subcritical $\lambda$ values.  The decay exponent $\alpha$ extrapolates to zero
at $\lambda = 2.760(2) \simeq \lambda_c$, as shown in the inset.  We note that the average is over {\it all} realizations,
including those that become inactive before the maximum time of $2 \times 10^6$ MCs.
If we instead restrict the average of $\rho(t)$ to trials that survive to time $t$ (or greater),
the result is a stretched exponential, $\rho(t) \sim \exp[-C t^\beta]$, where $C$ is a constant and the exponent
$\beta$ varies with $\lambda$.  For $\lambda = 2.70$, for example, $\beta \simeq 0.25$.

\subsection*{Threshold model simulations}

\begin{figure}[h]
\includegraphics[height=5.5cm]{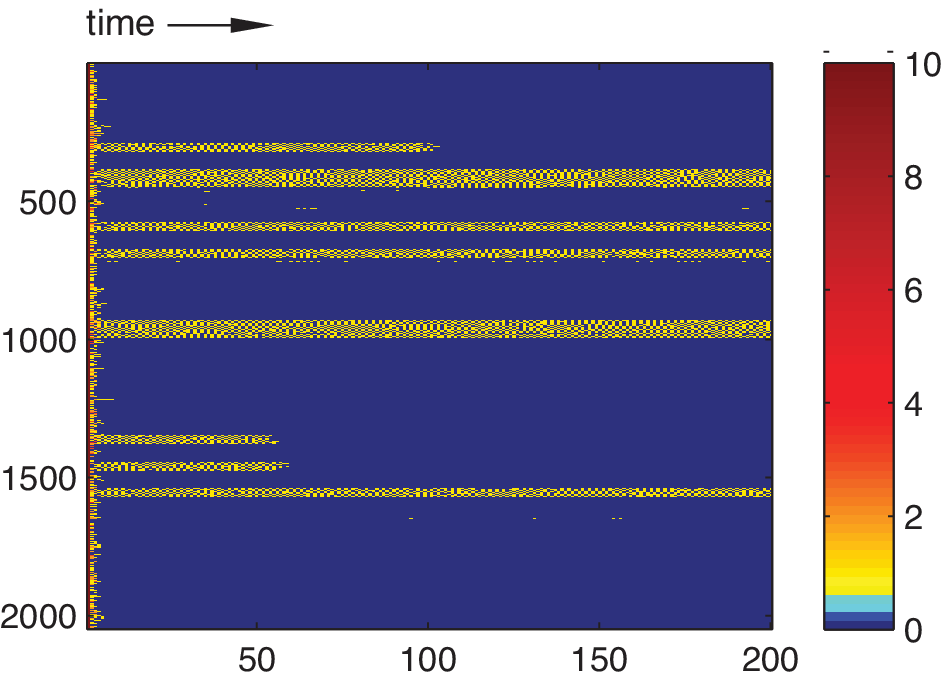}
\caption{\label{kep} 
Spatio-temporal evolution of the density of the $m=6$ threshold model,
proportional to the color coding on the right. The simulation is 
started from an active state, with $\nu = 0.9$ and $\lambda = 1$.
}
\end{figure}

Threshold models represent an attempt to capture the activation threshold 
of real neurons, by requiring at least $m$ active neighbors associated with 
incoming links to activate a node with probability $\lambda$. In case
of active nodes spontanous deactivation occurs with probability $\nu$
written in the reaction-diffusion model notation as:
\begin{description}
\item[(a)] $m A \to (m+1) A$ with probability $\lambda$,
\item[(b)] $A \to \emptyset $ with probability $\nu$.
\end{description}
We use SCA updating, as in \cite{KH}.
Since there is no spatial anisotropy (which might generate activity currents), 
we can assume that the SCA follow the same asymptotic dynamics as the 
corresponding model with random sequential updating. Thanks to the 
synchronous updates, we can speed up the simulations by $\sim n_p$ times 
by distributing the nodes among $n_p$ multiprocessor cores; by
swapping the random number generation on parallel running graphics 
cards we obtain a further reduction of $50\%$ decrease in the simulation times.
The spatio-temporal evolution of the HMN2d network with $m=6$, 
in a single network realization, starting from a fully active state, 
is shown in Figure~\ref{kep}.
After a sharp initial drop in activity, due to spontaneous deactivation 
of nodes with few neighbors, one observes domains (modules) on which 
activity survives for a long time, suggesting rare region effects.

\subsubsection*{Results for threshold models}

We begin by discussing the $m=2$ case, for which extensive 
simulations are performed. We generate networks with
average degree $\langle k\rangle = 24$ and $s=6$.
Note that in this case values of $\langle k\rangle$ higher than the 
percolation threshold are needed to avoid modules having separated activities.
As before, we averaged over hundreds of independent random networks
and thousands of independent realizations.  We followed the density
up to $t_{max}=10^5$ MCs, starting from a fully active initial condition.

\begin{figure}[h]
\includegraphics[height=5.5cm]{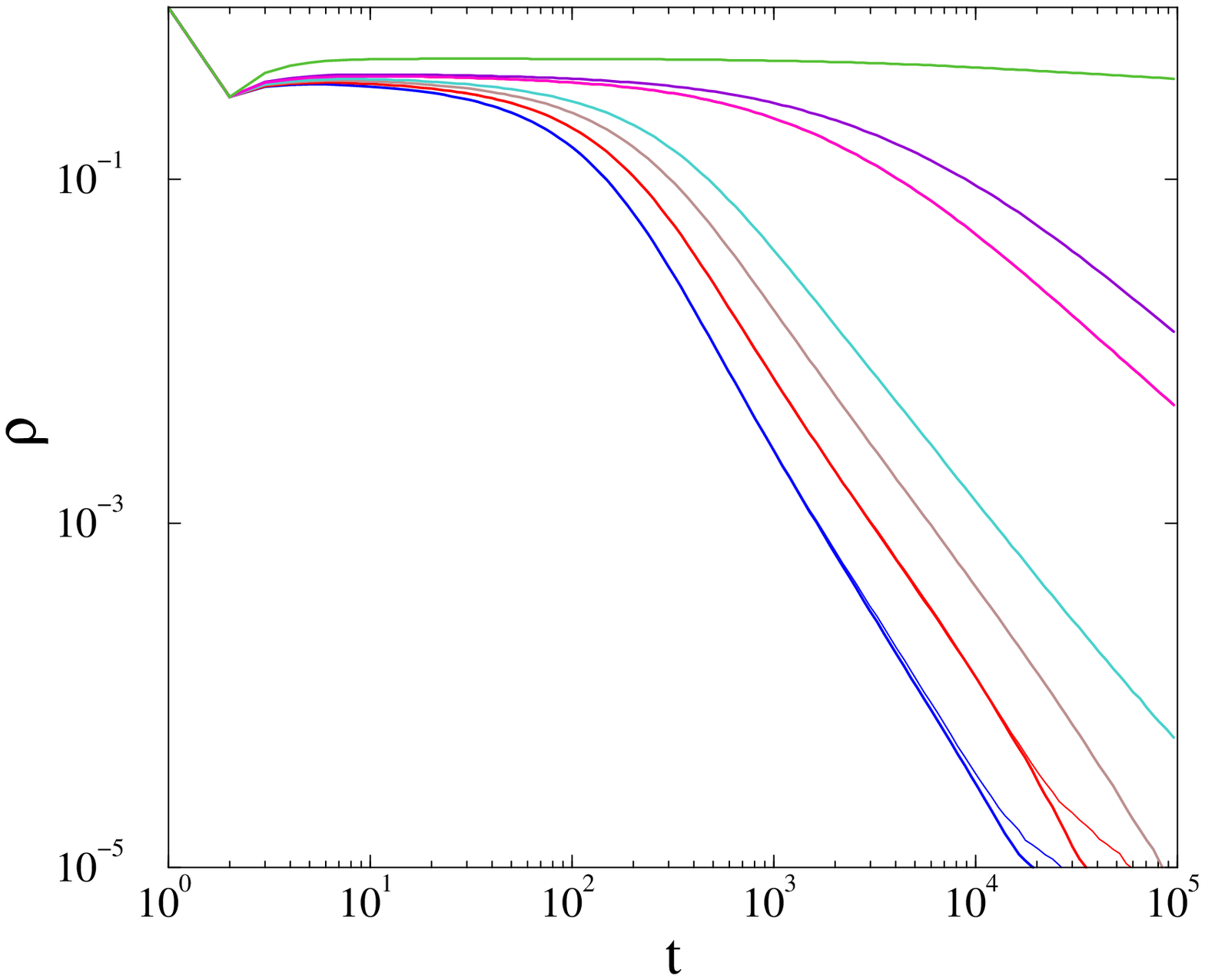}
\caption{\label{fss-k2}
Decay of activity in the $m = 2$ threshold model with $s = 6$, and 
$\langle k\rangle = 24$. The curves correspond to branching rates: 
$\lambda = 0.65, 0.66, 0.67, 0.68, 0.83$ (lower to upper) and $\nu =0.7$ fixed.
Levels: $l_{max}=8,9$ (thin, thick lines).
Size-independent power-laws, reflecting a GP are observed.}
\end{figure}

In this case the mean activity density decays more rapidly than in mean-field 
theory, and is size-independent (see Figure~\ref{fss-k2}). For large branching
probabilities, $\langle\rho(t)\rangle$ seems to take a constant value,
suggesting an active steady state, but at late times some deviation
is observed, possibly due to finite-size effects.

Homogeneous triplet creation models (i.e., $m=3$) are expected to exhibit
a mean-field-like {\it discontinuous} phase transition in two or more 
dimensions(see for example \cite{TCM}). 
Disorder induces rounding effects, producing continuous phase 
transitions or GPs \cite{round,round2}.  
We study a threshold model with $m=3$ and $s=6$, using $\langle k\rangle = 36$.
Our results are similar to those for $m=2$: non-universal
power-laws, again suggesting a GP.

\section*{Quenched Mean-field approximation for SIS}

Heterogeneous mean-field theory provides a good description of network models
when fluctuations are irrelevant. This approximation is attractive because
is can be solved analytically in many cases; the results
agree with simulation \cite{boguna09:_langev,Fer14}.
This analysis treats nodes with different degrees as distinct, but finally
averages over all degree values, providing a homogeneous solution for the order parameter.
To describe the effects of quasi-static heterogeneities in a more precise
way, the so-called Quenched Mean-Field (QMF) approximation was introduced
\cite{CWW08,GDOM12}. For SIS models this leads to a spectral
analysis of the adjacency matrix $A_{ij}$ of the network.
The susceptible-infected-susceptible (SIS) \cite{SIS} model is similar to
the CP, except that branching rates are not normalized by $k$, leading to
symmetric governing equations. A rate equation for the SIS model can be set
up for the vector of activity probabilities $\rho_i(t)$ of node $i$ at time $t$:
\begin{equation}
\label{qmfsis}
\frac{d\rho_i(t)}{dt} = -\rho_i(t) + \lambda (1-\rho_i(t))\sum_{j=1}^N A_{ij}w_{ij} \rho_j(t)~.
\end{equation}
Here the $w_{ij}=w_{ji}$ are weights attributed to the edges.
For large times the SIS model evolves to a steady state, with an order
parameter $\rho \equiv \av{\rho_{i}}$.
Since this equation is symmetric under the exchange $i \leftrightarrow j$, 
a spectral decomposition can be performed on a basis of orthonormal 
eigenvectors $\mbox{\boldmath$e$}(y)$.
\begin{equation}
\rho_i=\sum_{\Lambda} c(\Lambda) e_{i} (\Lambda).
\label{exp}
\end{equation}
Non-negativity of the matrix $B_{ij} \equiv A_{ij}w_{ij}$ assures a
unique, real, non-negative largest eigenvalue $y_{M}$.

In the long-time limit the system evolves into a steady state and
we can express the solution via $B_{ij}$ as
\begin{equation}
\rho_{i}=
\frac{\lambda\sum_{j}B_{ij}\rho_{j}}{1+\lambda\sum_{j}B_{ij}\rho_{j}} \ .
\label{SIS2}
\end{equation}
Stability analysis shows that $\rho_{i} > 0$ above a threshold $\lambda_{c}$,
with an order parameter $\rho \equiv \av{\rho_{i}}$.
In the eigenvector basis equation~(\ref{SIS2}) can be expanded by the coefficients
$c(\Lambda)$ as
\begin{equation}
c(\Lambda)=\lambda \sum_{\Lambda'} \Lambda' c(\Lambda') \sum_{i=1}^N \frac{e_i (\Lambda) e_i (\Lambda')}{1+\lambda \sum_{\widetilde{\Lambda}} \widetilde{\Lambda} c(\widetilde{\Lambda})e_i (\widetilde{\Lambda})}
\label{SIS4}
\end{equation}
and near the threshold we can express
$\rho_i \sim c_i(\Lambda_1) e_i(\Lambda_1)$ with the principal eigenvector.
In the QMF approximation $1/\lambda_c = y_{1}$ and
the order parameter can be approximated by the eigenvectors
of the largest eigenvalues
\begin{equation}
\rho(\lambda) \approx a_1 \Delta + a_2 \Delta^2 + ... \ ,
\end{equation}
where $\Delta = \lambda \Lambda_{1}{-}1 {\ll} 1$ with the coefficients
\begin{equation}
a_j = \sum_{i=1}^N e_i(\Lambda_j)/[N \sum_{i=1}^N e_i^3 (\Lambda_j)].
\label{epsilon}
\end{equation}

As proposed in \cite{GDOM12}, and tested on several SIS network models
\cite{basiscikk}, localization in the active steady
state can be quantified by calculating the inverse participation
ratio (IPR) of the principal eigenvector, related to the eigenvector of
the largest eigenvalue $\mbox{\boldmath$e$}(y_1)$ of the adjacency matrix as
\begin{equation} \label{defIPR}
I(N) \equiv \sum_{i=1}^{N} e_{i}^{4}(y_1) \ .
\end{equation}
This quantity vanishes $\sim 1/N$ in the case of homogeneous eigenvector
components, but remains finite as $N \to \infty$ if activity is concentrated
on a finite number of nodes.
Numerical evidence has also been provided for the relation of {\it localization
to RR effects, slowing down the dynamics to follow power-laws} 
\cite{basiscikk}.

\begin{figure}[h]
\includegraphics[height=5.5cm]{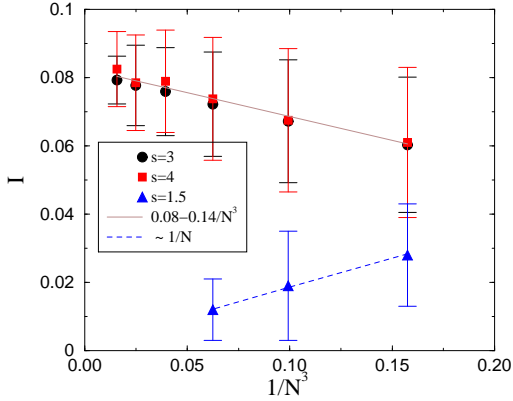}
\caption{\label{IPR}Finite size scaling of the inverse participation
ratio in weakly coupled HMN2d models, with maximum levels
$l_{max}=4,5,..9$. The $s=3$ (bullets) and $s=4$ (boxes) results suggest
localization (finite $I$) in the infinite-size limit. Lines are power-law
fits to the data. For $s=1.5$, corresponding to the symmetrized,
small-world network (model-6 of \cite{KH}) no evidence of localization is seen.}
\end{figure}

We study localization of the SIS in the HMN2d models introduced in
the previous section*s. We analyze the eigenvectors of $B_{ij}$ for
$b=2$ ($\langle k\rangle \simeq 4$), varying $s$, using system sizes
ranging from $N=256$ to $N=262144$.
In particular, we performed a FSS scaling study of the IPR in these systems.
First we determined the behavior on the small world network of
\cite{KH} corresponding to $s=3/2$. As one can see in Figure~\ref{IPR},
the localization at small sizes disappears as $I(N) \sim 1/N^3$.
By contrast, for $s=3$ and $s=4$, the graphs are finite-dimensional, and
the IPR increases with $N$, tending to a finite limiting value, suggesting localization.

\begin{figure}[h]
\includegraphics[height=5.5cm]{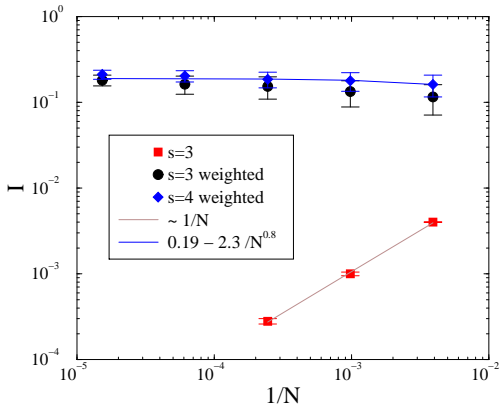}
\caption{\label{IPRW}Finite size scaling of the inverse participation
ratio in HMN2d models with higher average degree
($\langle k\rangle \simeq 50$) for maximum levels: $l_{max}=4,5,6,7,8$.
Bullets: $s=3$ with uniform randomly distributed weights; boxes: without
weights. Diamonds: $s=4$ with randomly distributed weights.
Lines show power-law fits to the data. In the unweighted case,
no localization effect can be seen, and $I$ decays linearly with $N$.}
\end{figure}

One might question the relevance of network models with small
connectivity to mammalian brains, in which
$\langle k\rangle$ is on the order of $10^3$. To answer this we study
$s \le 4$ models with higher connectivity, i.e., $\langle k\rangle \simeq 50$.
As Figure~\ref{IPRW} shows, the sign of localization, which is weak but nonzero for
$\langle k\rangle = 4$, now disappears.
Next, we add random weights $w_{ij}$, distributed uniformly over the interval $(0,1)$,
to the links of the networks. A consequence of this
explicit disorder is localization even in highly
connected networks, as shown in Figure~\ref{IPRW}.
This result is in agreement with the expectation of limited sustained activity
in brain networks \cite{KH}, meaning that the link disorder
prevents over-excitation of a network of high connectivity.

\begin{figure}[h]
\includegraphics[height=5.5cm]{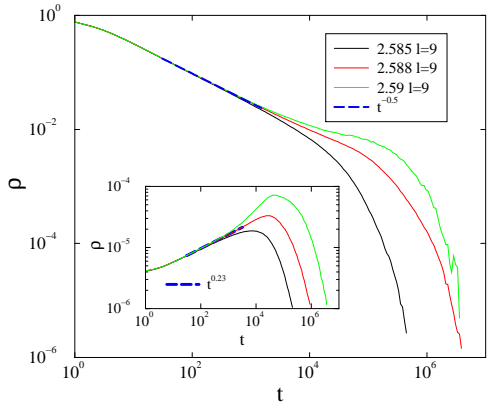}
\caption{\label{o.hiw6-s4-k50} Evolution of the activity in the
weighted CP on HMN2d networks with $s=4$, $l=9$, and $k=\langle 50\rangle$, 
for $\lambda = 2.585$, $2.588$, and $2.59$ (lower to upper). Main plot: 
decay in case of a fully active initial state. Inset: 
growth of activity starting from a single active node for the same 
values of $\lambda$.
Dashed lines are power-law fits for $\lambda_c\simeq 2.588$.}
\end{figure}

Localization suggests rare-region effects, thus a dynamic GP.
Nevertheless, simulations of the CP on such networks do not show extended
regions of power-laws for $s \le4$, but rather a nontrivial (non-mean-field)
continuous phase transition (Figure~\ref{o.hiw6-s4-k50}).
Decay simulations for $l=9$ yield
$\rho(t) \simeq t^{-0.50(1)}$ at $\lambda_c\simeq 2.588(1)$, albeit
with a cutoff due to the finite system size.
This agrees with our result for 1D BB networks with $s=3$ \cite{GPCNlong}.
Spreading simulations starting from single, randomly placed seed result
in the survival probability scaling at this critical
point: $P(t) \simeq t^{-0.50(1)}$ (that is, the symmetry $\alpha=\delta$ holds to within uncertainty).
Here the density increases initially as $\rho(t) \simeq t^{0.23(1)}$.
One can understand these results, noting that the localization values are rather small,
$I \sim 0.08$, so that RR effects are too weak to generate observable
GP effects in the dynamics.

Naturally, for $s > 4$ networks, where already the topological disorder
is strong enough to create a GP, inhomogeneities in the interaction
strengths amplify the RR effects. This induces GPs even for larger values of
$\langle k\rangle$, as well as in the threshold models.
Further studies of disorder effects are under way.

\section*{Conclusions}

We investigate the dynamical behavior of spreading models on hierarchical 
modular networks embedded in two-dimensional space, 
with long links whose probability decays as a power-law with distance.
This corresponds to an exponentially decreasing connection probability, 
as a function of the level in the hierarchical construction. 
The aim of this study is to understand the effects of intrinsic 
and topological disorder, in particular, extended critical regions 
in the parameter space, without any (self) tuning mechanism.

If we eliminate the underlying lattice structure we observe power-law
dynamics for networks near the percolation threshold, when the effective
dimension is finite. However, size-independent power-laws, corresponding
to GPs, are only seen if we have directed links.
Since connectivity of these structures is not guaranteed, we also study random
hierarchical trees, with full connectivity. GPs are observed upon
averaging over many independent network realizations. The relation to brain
networks can be envisaged in the large-size limit, if we regard independent
realizations as (almost decoupled) sub-modules of the entire brain.

When we ensure connectivity via short edges on the lowest level, we
find GPs for rapidly decaying long-link probabilities. Both of
these network assumptions are in accordance with empirical brain
networks. Above the GP, at the critical point, we find mean-field-like
decay of activity, in agreement with results on the CP with quenched disorder 
on higher-dimensional, regular lattices.  We have also shown that bursty 
behavior arises naturally in the GPs, due to autocorrelations which 
decay via a power-law.

We perform a quenched mean-field study of the SIS model on
these networks; in the SIS, nodes are connected
symmetrically. Finite size scaling of the inverse participation ratio
suggests localization on large-world networks and de-localization on small
world structures. However, when we add intrinsic weight disorder,
localization can be seen even on small-world networks. 
Weight disorder is again to be expected in real brain networks, 
since the strength of couplings varies over many orders of magnitude.
Despite this, we saw no GP effects in the dynamics of
weighted CPs with $s=4$. Instead, we find a nontrivial critical 
scaling, as has already been observed in other networks.
The possibility of a narrow GP in this case is an open issue.

In conclusion, we believe our synthetic HMN2ds are closer to experimental
brain networks than previously proposed models, and find numerical evidence for
GPs in extended phases in simple models with spreading dynamics.
Although we eliminate any self-tuning mechanisms, we
still find nontrivial slow dynamics as well as localization of activity, which
is crucial for understanding real brain network data. 

\section*{Acknowledgments}

We thank R. Juh\'asz and C. C. Hilgetag for useful discussions.
Support from the Hungarian research fund OTKA (Grant No. K109577)
and the European Social Fund through project FuturICT.hu (grant no.:
TAMOP-4.2.2.C-11/1/KONV-2012-0013) is acknowledged.
R.D. is supported in part by CNPq, Brazil.
The SCA simulations are accelerated by grapics cards donated
by NVIDIA within the professor partnership programme.
G\'eza \'Odor thanks the Physics Department at UFMG, where part
of this work was done, for its hospitality.

\section*{Author contributions statement}
G\'eza \'O. wrote, ran and analysed the CP and SIS simulations
on HMN2d-s and performed the QMF analysis. R. D. wrote, ran and analyzed
the HMT simulations. Gergely \'O. wrote and ran the HMN2 threshold
model simulations and dimension measurements on them.
G\'eza \'O. and R. D. wrote the main manuscript text.
G\'eza \'O. prepared figures 1-3, 6-9, 13-16,
R. D. prepared figures 4,5,10,11,17-20. Gergely \'O. prepared figure 12.
All authors reviewed the manuscript.

\section*{Additional information}
The authors declare no competing financial interests.

\section*{Supplementary information}

Although the set of nodes is isomorphic to a square lattice of
$N= 2^L \! \times 2^L$ sites, we shall
label the nodes not by their Cartesian coordinates but rather using a
base-4 notation.  At level $l$, the full set of $4^l$ nodes is divided into
four quadrants labeled 0, 1, 2, and 3, proceeding counterclockwise from the
lower left.  At level $l-1$, each of the quadrants is similarly divided
into four subquadrants, labeled in the same manner.
Division into subquadrants continues all the way down to level 1, in which
the elements are individual nodes.
A given site can be specified by the labels of the quadrant, subquadrant, etc.
to which it belongs.
Denoting the labels at levels 1, 2,..., L, by $g_1, ..., g_L$, the address of
a site is $n = \sum_{l=1}^L g_l \, 4^{l-1}$ (see Fig. S17).
\vspace{1em}

\captionsetup[figure]{name=Supplementary Figure S}
\begin{figure}[!h]
\includegraphics[width=5cm]{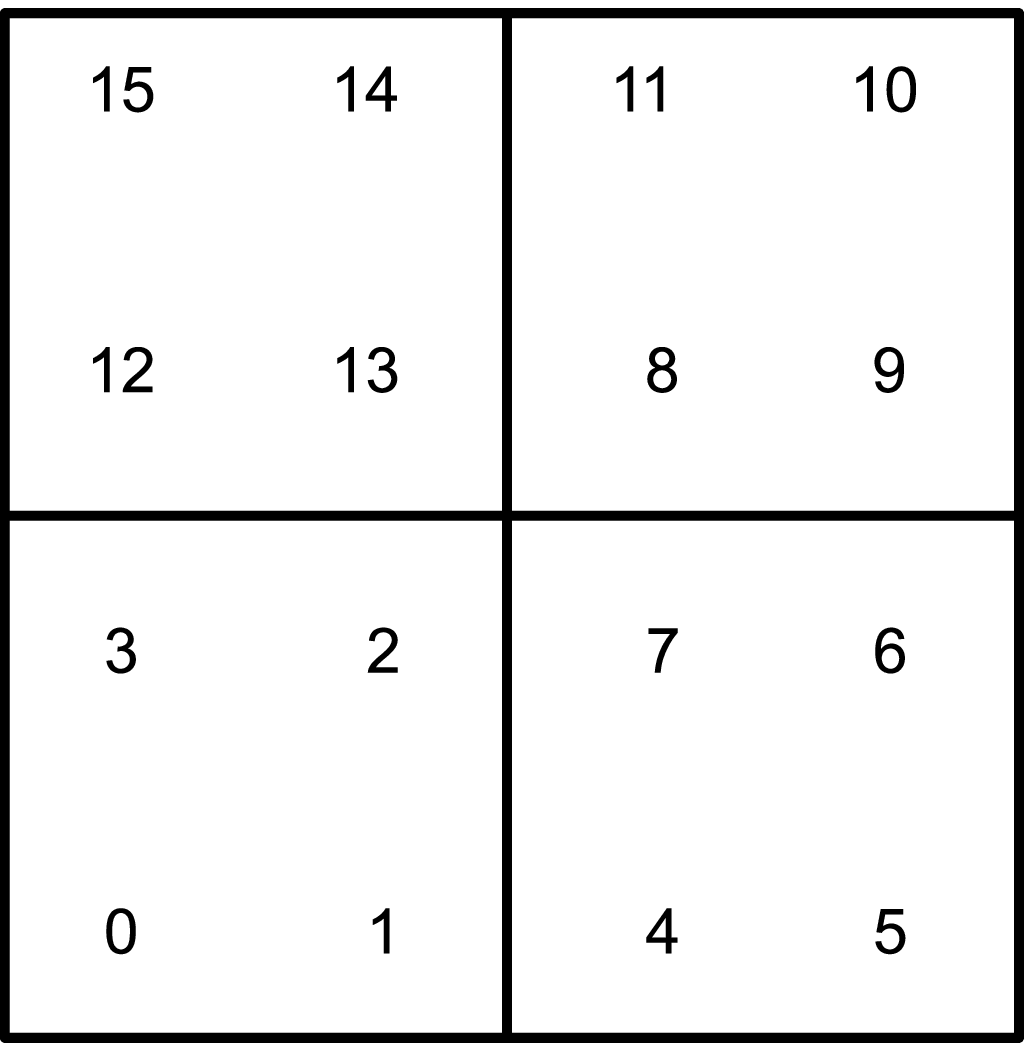}
\caption{Definition of node labels for a network with $L=2$} \label{module}
\end{figure}

Consider a set of four nodes, 0, 1, 2, and 3.  There are six possible links between them: (0,1), (0,2), (0,3),
(1,2), (1,3), and (2,3).  A tree with four nodes has exactly three links.  Of the $\binom{6} {3} = 20$ subsets of
three links, sixteen yield a connected tree [for example: \{(0,1), (1,2), (2,3)\}],
while four correspond to a cycle of three nodes, leaving the fourth node isolated [example: \{(1,2), (1,3), (2,3)\}].
To construct a tree of four nodes, we choose one of the sixteen link sets ${\cal B}_i$ at random.

A hierarchical tree is constructed by first linking the four quadrants via three edges.  The link set
${\cal B}_i$ is chosen at random.  Then, for each link, we choose sites at random within each of the
two quadrants connected by the link, to serve as the connected nodes.  Now we repeat this process
within each of the subquadrants, and so on, until we reach the basic modules of four sites.  The latter
are again connected by sets of three links, chosen independently from the basic collection of sixteen
link sets.  At the end, each basic module is connected internally, and to a module at the next level, etc.,
so that we have a connected graph of $N$ nodes and $N-1$ edges.  Although we have chosen to begin the construction
at level $L$, we could equally have begun at level 1, or some intermediate level: the choices of link sets
and nodes at the various levels are mutually independent.  These steps are illustrated in the Figs. S18, S19, S20.
\vspace{1em}

\begin{figure}[!h]
\includegraphics[width=5cm]{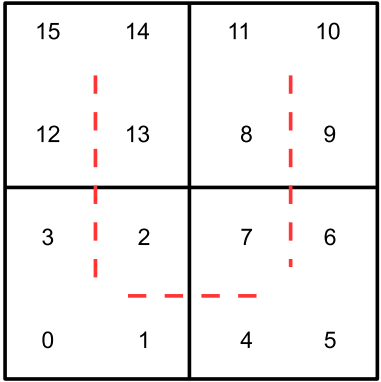}
\caption{First step in constructing a network with $L=2$: define the link set at level 2.} \label{module1}
\end{figure}
\vspace{1em}

\begin{figure}[!h]
\includegraphics[width=5cm]{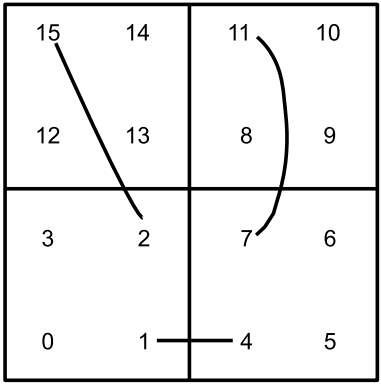}
\caption{Second step: choose nodes consistent with link set at level 2.
At the next step (not shown) we choose link sets for each of the four-site modules.} \label{module2}
\end{figure}
\vspace{2em}

\begin{figure}[!h]
\includegraphics[width=5cm]{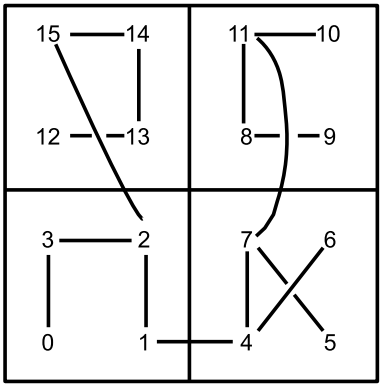}
\caption{Final step: choose nodes at level 1 consistent with the corresponding link sets.} \label{module3}
\end{figure}

At level 1 the number of links per node is
3/4; at level $j$, there are $4^{L-j}$ blocks connected via $4^{L-j}-1$ links.  The number of links per node
at this level is therefore $(4^{L-j}-1)/4^L \simeq 1/4^j$.  Let $p_j$ denote the probability that a randomly chosen
edge linking blocks at level $j$ be present.  At level 1, links connect nodes within four-site modules.
Since there are six possible links, and since just three are present within each module,
we have $p_1 = 1/2$.  At level 2, a link connects nodes within a 16-node module; at this level a node is linked to
one of the 12 nodes outside its basic 4-site module.  There are $(16 \cdot 12)/2 = 96$ possible links, and since only
three are present, $p_2 = 1/32$.  At level $j$, an edge links nodes within a module containing $4^j$ nodes.
The number of possible links at this level is $4^j (3\cdot 4^{j-1})/2$, so that $p_j = 1/2^{4j-3}$, and, in general,
$p_j/p_{j-1} = 1/16$.  Since a tree represents a minimally connected structure, $p_j$ cannot decay faster than
this rate, in any connected hierarchical network based on four-node modules.

Although our principal interest is in random trees, as described above, we may also consider {\it regular} trees;
these are constructed by using the same link set at each level.  The resulting structure turns out to have infinite
topological dimension, as discussed in Section "Dimension measurements".

\end{document}